\documentclass[12pt]{article}
\input epsf.sty
\newcommand{\tl}{\wt\lambda}

\newcommand{\HH}{{\cal H}}

\newcommand{\OO}{{\cal O}}

\newcommand{\wt}{\widetilde}

\newcommand{\TT}{{\cal T}}

\newcommand{\be}{\begin{equation}}
\newcommand{\ee}{\end{equation}}
\newcommand{\ben}{\begin{eqnarray}\displaystyle}
\newcommand{\een}{\end{eqnarray}}
\newcommand{\refb}[1]{(\ref{#1})}
\newcommand{\p}{\partial}

\begin{document}
{}~
\hfill\vbox{\hbox{hep-th/0204143}
}\break

\vskip .6cm

\centerline{\Large \bf
Field Theory of Tachyon Matter}

\medskip

\vspace*{4.0ex}

\centerline{\large \rm 
Ashoke Sen }

\vspace*{4.0ex}

\centerline{\large \it Harish-Chandra Research
Institute}

\centerline{\large \it  Chhatnag Road, Jhusi,
Allahabad 211019, INDIA}

\centerline {and}
\centerline{\large \it Department of Physics, Penn State University}

\centerline{\large \it University Park,
PA 16802, USA}

\centerline{E-mail: asen@thwgs.cern.ch, sen@mri.ernet.in}

\vspace*{5.0ex}

\centerline{\bf Abstract} \bigskip 

We propose a field theory for describing the tachyon on a brane-antibrane
system near the minimum of the potential. This field theory realizes two
known properties of the tachyon effective action: 1)  absence of
plane-wave solutions around the minimum, and 2) exponential fall off of
the pressure at late time as the tachyon field evolves from any spatially
homogeneous initial configuration towards the minimum of the potential.
Classical solutions in this field theory include non-relativistic matter 
with arbitrary spatial distribution of energy.

\vfill \eject

\baselineskip=16pt



Dynamics of the tachyon on D-brane anti-D-brane system or unstable 
D-branes in superstring theory and bosonic string 
theory have been investigated during the last 
few years. Of the various 
known 
properties of the classical tachyon effective action,\footnote{Although we 
shall use 
the language of tachyon effective action obtained by integrating out the 
other open string modes, we should note that many of these 
results are derived using
the technology of full string field theory, containing all the 
infinite number of fields describing the dynamics of the open string.} two 
specific properties 
which specify the behaviour of the tachyon effective action around the 
minimum of the potential are as follows:
\begin{enumerate}
\item Since the minimum of the potential describes a configuration 
where there are no D-branes\cite{origin}, around this minimum 
there are no physical 
open string excitations. Translated to a property of the tree level 
tachyon effective action, this implies that there are no plane wave 
solutions to the linearized equations of motion around the minimum of the 
tachyon potential.

\item If we let the tachyon roll beginning with any spatially homogeneous 
initial configuration\cite{0202210}, it evolves asymptotically 
towards its minimum instead of oscillating about the 
minimum\cite{0203211,0203265}. The 
total 
energy density is conserved during the evolution but the pressure evolves 
to zero. The information about the tachyon effective action near the 
minimum of the potential is encoded in the late time behaviour of the 
pressure. This is given by\cite{0203211,0203265}\footnote{Since we are 
interested in studying 
tree level properties of the tachyon potential, we shall ignore the 
backreaction of gravity and other closed string fields on the evolution of 
the tachyon. Once the form of the action is derived by this method, one 
can of course couple it to background gravitational field and study the 
evolution of the coupled system. These effects have been studied in 
refs.\cite{cosmo}. For some 
earlier attempts at the study of time dependent solutions for the tachyon 
field see \cite{earl}.}
\be \label{e1}
p = -K e^{-\alpha x^0}\, ,
\ee
where $K$ is an irrelevant normalization constant which can be changed by 
shifting the origin of the time coordinate $x^0$, and $\alpha$ is given 
(in the $\alpha'=1$ unit) by
\ben \label{e2}
\alpha &=& 1 \qquad \hbox{for bosonic string theory} \nonumber \\
&=& \sqrt 2 \qquad \hbox{for superstring theory}\, .
\een

\end{enumerate}

Several field theory models have been proposed for realising the first 
property. These models are of two kind.\footnote{Here we are stating the 
results in terms of properly redefined field $\phi$ for which the two 
derivative 
term has the standard form $-{1\over 2}\eta^{\mu\nu}\p_\mu\phi 
\p_\nu\phi$.}
In the first kind\cite{padic} the 
tachyon 
potential is non-singular at the minimum of the potential but the absence 
of physical plane-wave solutions has its origin in the structure of the 
kinetic term (which typically involves higher derivative 
terms). This possibility seems to be realized\cite{sft} in the explicit 
analysis of the string field theory based on $*$-product 
interaction\cite{cubic}.
In the second kind of field theory models\cite{zwie} the tachyon potential 
is singular at the minimum. In particular the second derivative of the 
potential blows up and hence the tachyon mass becomes infinite. Analysis 
based on boundary string field theory\cite{bsft} seems to support this 
view\cite{bsfttach}. Of 
course these results are not mutually incompatible, since an effective 
field theory of the first kind may be turned into one of second kind and 
vice versa by a suitable field redefinition that involves (infinite number 
of) derivatives of the tachyon field.

A field theory model that realises the evolution of the system to a 
pressureless gas was proposed in \cite{0203265} following earlier 
proposals in 
\cite{effective,0009061}. The action was taken to be
\be \label{ey1}
S= - \int d^{p+1} x \, V(T) \sqrt{-\det A}\, ,
\ee
where
\be \label{ey2}
A_{\mu\nu} = \eta_{\mu\nu} + \p_\mu T \p_\nu T\, .
\ee
However in this proposal the tachyon potential $V(T)$ was left 
undetermined.  The only property of $V(T)$ that was determined is that
the minimum of $V(T)$ where $V(T)=0$ should be at  
$T=\infty$. 

In this paper we shall show that by taking $V(T)\propto e^{-\alpha T/2}$ 
for large $T$ in the action \refb{ey1}, we can satisfy both the 
requirements given above: the absence of plane-wave solutions around the 
tachyon vacuum, and the exponential fall off of the pressure for large 
$x^0$. The constant of proportionality can be absorbed into a constant 
shift of 
$T$.
Thus the proposed tachyon effective action for large $T$ on an 
unstable 
D-$p$-brane system  is
\be \label{ez1}
S= - \int d^{p+1} x \, e^{-\alpha T/2} \sqrt{-\det A}
= - \int d^{p+1} x \, e^{-\alpha T/2} \, \sqrt{1 + \eta^{\mu\nu}\p_\mu T 
\p_\nu T}\, ,
\ee
where $A_{\mu\nu}$ is given by eq.\refb{ey2}. We expect this to be a valid 
approximation when the second and higher order derivatives 
of $T$ 
are small. As we shall see, the tachyon matter does satisfy this 
condition.

On a brane-antibrane system 
where $T$ is a complex field, the tachyon potential is given by 
$e^{-\alpha |T|/2}$. In this case the action \refb{ez1} can be regarded as 
the restriction of the full action to the case $\Im(T)=0$.
We shall carry out our analysis for this restricted field configuration.
The imaginary part of the tachyon, which 
can be regarded as the Goldstone mode associated to the broken $U(1)$ 
phase symmetry, is absorbed by the U(1) gauge field due to Higgs 
mechanism, and is expected to decouple from the dynamics near the minimum 
of the potential.

We shall first verify that the action \refb{ez1} produces the correct 
large 
$x^0$ behaviour of the pressure. For this analysis we can restrict to 
spatially homogeneous, time dependent field configurations. As discussed 
in \cite{0009061,0203265}, for such configurations the conserved energy 
density is 
given 
by:
\be \label{e2b}
T_{00} = e^{-\alpha T/2} (1-(\p_0T)^2)^{-1/2}\, .
\ee
Since $T_{00}$ is conserved, we see that for any given $T_{00}$,
as $T\to\infty$, $\p_0 T\to 1$. In particular, for large $x^0$ the 
solution has the 
form:\footnote{Note that since for this configuration $\p_0^n T$ for $n>1$ 
falls off exponentially, the contribution from possible higher derivative 
corrections like $(\p_\mu\p^\mu T)^2$ inside the square root on the 
right hand side of \refb{ez1} will be suppressed.}
\be \label{e2a}
T = x^0 +  C e^{-\alpha x^0} + \OO(e^{-2\alpha x^0}) \, .
\ee
In order to see that \refb{e2a} gives the correct form of the solution we 
simply need to note that the leading contribution to $T_{00}$ 
computed 
from this configuration
remains constant in time:
\be \label{e3}
T_{00} \simeq {1\over 
\sqrt{2\alpha C}}\, .
\ee
The pressure associated with this configuration is given 
by\cite{0009061,0203265}:
\be \label{e4}
p= -e^{-\alpha T/2}(1-(\p_0T)^2)^{1/2} \simeq -\sqrt{2\alpha C} 
e^{-\alpha x^0}\, .
\ee
This is in precise agreement with \refb{e1}.

Next we shall demonstrate the absence of plane-wave solutions. 
First let us expand the action in powers of derivatives of $T$ up to terms 
containing at most two powers of derivatives. This gives
\be \label{e6}
S= - \int d^{p+1} x \, e^{-\alpha T/2} (1 +{1\over 2}\, \eta^{\mu\nu} 
\p_\mu T \p_\nu T
+ \ldots)\, ,
\ee
where $\ldots$ denotes terms with higher powers of derivatives.
Defining 
\be \label{e7}
\phi = e^{-\alpha T/4}\, ,
\ee
we get 
\be \label{e8}
S=  \int d^{p+1} x \, {16\over \alpha^2} (-{1\over 2}\eta^{\mu\nu} 
\p_\mu\phi \p_\nu\phi
-{\alpha^2\over 16} \phi^2 + \ldots)\, .
\ee
The minimum of the potential is at $\phi=0$. If we ignore the higher 
derivative terms then the quantization of the action seems to lead to a 
scalar particle of mass $\alpha/2\sqrt2$. This is reflected in the fact 
that the equations of motion without the higher derivative terms contain a 
plane-wave solution of the form:
\be \label{e9}
\phi = a \, e^{ik_\mu x^\mu}\, ,
\ee
with $-\eta^{\mu\nu} k_\mu k_\nu= \alpha^2/8$ and $a$ an arbitrary 
constant.

We shall now show that when we take into account the effect of the higher 
derivative terms, \refb{e9} ceases to be a solution of the equations of 
motion for any $k_\mu$.\footnote{Note that \refb{e9} describes a complex 
solution. Since the full equations of motion are homogeneous of 
degree one, but not linear, even 
if we had a solution of the form \refb{e9}, it is not clear how we could 
construct a real solution since we cannot superpose the solutions. We 
shall show however that even this complex solution does not exist.}
To show this we note that the full action 
\refb{ez1} written in 
terms of $\phi$ defined in \refb{e7} takes the form:
\be \label{e10}
S= - \int d^{p+1} x \, \phi^2 \sqrt{1 + {16\over \alpha^2} 
\phi^{-2} \eta^{\mu\nu}\p_\mu\phi \p_\nu\phi
}\, .
\ee
The action is homogeneous of degree 2 in $\phi$. The equations of motion 
derived from this action is given by:
\be \label{e11}
-\eta^{\mu\nu}\p_\mu \bigg( {\p_\nu \phi \over 
\sqrt{1 + {16\over \alpha^2}
\phi^{-2} \eta^{\mu\nu}\p_\mu\phi \p_\nu\phi
}}\bigg) 
+ {\alpha^2\over 8} \, {\phi + {8 \over \alpha^2}
\phi^{-1} \eta^{\mu\nu}\p_\mu\phi \p_\nu\phi
\over \sqrt{1 
+ {16\over 
\alpha^2}
\phi^{-2} \eta^{\mu\nu}\p_\mu\phi \p_\nu\phi
}}=0\, .
\ee
Substituting \refb{e9} into \refb{e11} we get:
\be \label{e12}
{a\over \sqrt{1 - {16\over \alpha^2} \eta^{\mu\nu}k_\mu k_\nu
} } = 0\, .
\ee
Clearly this equation has no non-trivial solution for finite values of 
$k_\mu$. 
This establishes the absence of plane-wave solutions.\footnote{Note again 
that since for the configuration \refb{e9} $\p_\mu \p_\nu T$ vanishes, 
contribution to the equations of motion from possible higher derivative 
corrections like $(\p_\mu \p^\mu T)^2$ inside the square root on the right 
hand 
size of \refb{ez1} vanishes.} (Note that if we 
expand the left hand side of this equation in powers of $k_\mu$ and keep 
up to quadratic terms, we reproduce the mass-shell condition 
$-\eta^{\mu\nu} k_\mu k_\nu
= \alpha^2/8$ derived earlier.)

Absence of plane wave solutions does not imply absence of 
other classical solutions
however. We have already seen the existence of solutions with constant 
(but arbitrary) energy density. As emphasized in \cite{0009061} (see also 
\cite{0002223,older}) the correct excitations of the system can be found 
by 
working in 
the hamiltonian formalism. Defining the momentum conjugate to $T$ as:
\be \label{ex1}
\Pi(x) = {\delta S \over \delta (\p_0 T(x))}\, ,
\ee
we can construct the Hamiltonian $H$ following \cite{0009061}:
\be \label{ex2}
H = \int d^p x \HH, \qquad
\HH = T_{00} = \sqrt{\Pi^2 + e^{-\alpha T}} \, \sqrt{1 + \p_i T \p_i T}
\, .
\ee
For large $T$ we can ignore the $e^{-\alpha T}$ term, and the equations 
of motion take the form:
\be \label{ex4}
\p_0 \Pi(x)= -{\delta H\over \delta T(x)} = \p_j \bigg( \Pi(x)\, { \p_j T 
\over 
\sqrt{1 + \p_i T \p_i T} }\bigg)\, .
\ee
\be \label{ex5}
\p_0 T(x) = {\delta H \over \delta \Pi(x)} = \sqrt{1 + \p_i T \p_i T} \, ,
\ee
for $\Pi(x)> 0$.
Thus we can get a solution to the equations of motion by taking $\p_i T = 
\p_0 \Pi =0$, $\p_0 T = 1$. This gives
\be \label{ex6}
\Pi(x) = f(\vec x)\, , \qquad T(x) = x^0\,, 
\ee
where $f(\vec x)$ is any arbitrary function of the spatial coordinates. 
The energy density associated with such a solution is proportinal to 
$f(\vec x)$.
This shows that the system admits classical solutions with energy density 
which is time 
independent but has arbitrary dependence on the spatial coordinates.
Classically this energy density can be as low as we like. 
The ability to create configurations with arbitrarily low energy density 
can
be traced to the scale invariance:
\be \label{ex7}
T(x)\to \lambda^{-1} T(\lambda x), \qquad
\Pi(x)\to \lambda^{p+1} \Pi(\lambda x)\, ,
\ee
under which the Poisson brackets and the equations of motion 
remain unchanged, and $\HH(x) \to \lambda^{p+1} \HH(\lambda x)$.

Since the original theory is Lorentz invariant, it is clear that given a 
localized density of tachyon matter we should be able to boost it to an 
arbitrary velocity. Intuitively we also expect that is should be 
possible to construct configurations with different local velocities at 
different points. This can be seen to be the case by rewriting the 
equations of motion in a slightly different form. We define:
\be \label{eadd1}
u_\mu = \p_\mu T, \qquad \epsilon(x) = \Pi(x)/ \p_0 T(x)\, .
\ee
The equations of motion then take the suggestive form:
\be \label{eadd2}
\eta^{\mu\nu} u_\mu u_\nu = -1, \qquad \p_\mu (\epsilon(x) u^\mu) = 0\, .
\ee
The energy momentum tensor can be computed following \cite{0009061}. 
Expressed in terms of these new variables, $T_{\mu\nu}$ take the form: 
\be \label{eadd3}
T_{\mu\nu} = \epsilon(x) u_\mu u_\nu\, .
\ee
These are precisely the equations govering gradient flow of 
non-interacting dust, 
with $u_\mu$ interpreted as the local velocity vector. 
These equations are expected to be valid as long as $\p_\mu u_\nu$ is 
small in magnitude (in string units).

Although in our analysis we have ignored the coupling of the tachyon to 
various massless fields on the D-brane world-volume, these can be easily 
incorporated following \cite{as,effective,0009061,0002223}.
We can generalize \refb{ez1} to curved background space-time 
by 
replacing $\eta_{\mu\nu}$ in eq.\refb{ey2} by the closed string metric 
$g_{\mu\nu}$. 
If we also have anti-symmetric tensor field background then 
$\eta_{\mu\nu}$ should be replaced by $g_{\mu\nu}+B_{\mu\nu}$.
The dilaton 
$\phi$ couples via an overall factor of $e^{-\phi}$ 
multiplying the lagrangian density.
For non-BPS D-$p$-brane of type II string theories, another quantity of 
interest is the coupling of the tachyon to Ramond-Ramond (RR) field 
$C^{(p)}$. In 
particular it is known\cite{rrform} that the D-$p$-brane 
world-volume theory has 
a coupling:
\be \label{exy1}
\int d^{p+1} x \, f(T) \, dT \wedge C^{(p)}\, ,
\ee
where $f(T)$ is some function of the tachyon field $T$. We can determine 
the behaviour of $f(T)$ 
for large $T$ if we can find the source of the RR field that is generated 
by the rolling tachyon field\cite{0202210}. This can be computed from the 
boundary state 
associated with the solution. Although we have not performed a systematic 
analysis of the problem along the lines of \cite{0203211,0203265}, general 
arguments based on symmetry and other considerations lead us to guess 
that (in 
$\alpha'=1$ unit) the source for the RR 
$p$-form field is proportional to:
\be \label{exy2}
\sin(\tl\pi) \, \bigg[ {e^{x^0/\sqrt 2} \over 1 + \sin^2(\tl\pi)e^{\sqrt 2 
x^0} } - {e^{-x^0/\sqrt 2} \over 1 + \sin^2(\tl\pi)e^{-\sqrt 2
x^0} }\bigg] \, ,
\ee
where $\tl$ is the parameter labelling the total energy density of the 
system:
\be \label{exy3}
T_{00} = {\TT_p\over 2} \, (1 + \cos(2\tl\pi) )\, ,
\ee
$\TT_p$ being the tension of the non-BPS D-$p$-brane. Thus for large $x^0$ 
the 
source is proportional to $e^{-x^0/\sqrt 2}$. Since $T\simeq x^0$ for 
large $x^0$, this indicates that for large $T$, the coupling of the RR 
$p$-form field to the tachyon has the form:
\be \label{exy4}
\int d^{p+1} x \, e^{-T/\sqrt 2}\,  dT \wedge C^{(p)}\, ,
\ee
up to an overall normalization constant.

There are many issues which call for further investigation.
As we have seen, the effective field theory describing tachyon matter has 
classical solutions with localized energy densities.
It will be interesting to construct the two dimensional conformal field 
theories associated with these solutions 
along 
the lines of 
refs.\cite{0203211,0203265} 
where the conformal field theories associated with spatially homogeneous 
energy densities were constructed. (Of course, one possibility is to take 
an arbitrary spatial distribution of coincident D0-$\bar{\rm D}0$ pair and 
construct the rolling tachyon solution on each such pair.) Another issue 
of importance is the effect of quantum corrections on such background.

\medskip

{\bf Acknowledgement}:
I would like to thank L.~Kofman, J.~Maldacena and T.~Padmanabhan for 
useful
discussions.
This work was supported in part by a grant 
from the Eberly College 
of Science of the Penn State University.


\begin{thebibliography}{99}

\bibitem{origin}
A.~Sen,
JHEP {\bf 9808}, 010 (1998)
[arXiv:hep-th/9805019];
JHEP {\bf 9808}, 012 (1998)
[arXiv:hep-th/9805170];
JHEP {\bf 9912}, 027 (1999)
[arXiv:hep-th/9911116].

\bibitem{0202210}
M.~Gutperle and A.~Strominger,
``Spacelike branes,''
arXiv:hep-th/0202210.

\bibitem{0203211}
A.~Sen,
``Rolling Tachyon,''
arXiv:hep-th/0203211.

\bibitem{0203265}
A.~Sen,
``Tachyon matter,''
arXiv:hep-th/0203265.

\bibitem{cosmo}
G.~W.~Gibbons,
arXiv:hep-th/0204008;
M.~Fairbairn and M.~H.~Tytgat,
arXiv:hep-th/0204070;
S.~Mukohyama,
arXiv:hep-th/0204084;
A.~Feinstein,
arXiv:hep-th/0204140;
T.~Padmanabhan, arXiv:hep-th/0204150;
A.~Frolov, L.~Kofman and A.~A.~Starobinsky,
arXiv:hep-th/0204187;
D.~Choudhury, D.~Ghoshal, D.~P.~Jatkar and S.~Panda,
arXiv:hep-th/0204204;
X.~Li, J.~Hao and D.~Liu,
arXiv:hep-th/0204252;
G.~Shiu and I.~Wasserman,
arXiv:hep-th/0205003.




\bibitem{earl}
C.~Acatrinei and C.~Sochichiu,
arXiv:hep-th/0104263;
S.~H.~Alexander, Phys. \ Rev. \ D {\bf 65}, 023507 (2002)
[arXiv:hep-th/0105032];
A.~Mazumdar, S.~Panda and A.~Perez-Lorenzana,
Nucl.\ Phys.\ B {\bf 614}, 101 (2001)
[arXiv:hep-ph/0107058];
S.~Sarangi and S.~H.~Tye,
arXiv:hep-th/0204074.

\bibitem{padic}
L.~Brekke, P.~G.~Freund, M.~Olson and E.~Witten,
Nucl.\ Phys.\ B {\bf 302}, 365 (1988);
P.~H.~Frampton and Y.~Okada,
Phys.\ Rev.\ D {\bf 37}, 3077 (1988);
D.~Ghoshal and A.~Sen,
Nucl.\ Phys.\ B {\bf 584}, 300 (2000)
[arXiv:hep-th/0003278];
J.~A.~Minahan,
JHEP {\bf 0103}, 028 (2001)
[arXiv:hep-th/0102071].

\bibitem{sft}
V.~A.~Kostelecky and S.~Samuel,
Nucl.\ Phys.\ B {\bf 336}, 263 (1990);
A.~Sen and B.~Zwiebach,
JHEP {\bf 0010}, 009 (2000)
[arXiv:hep-th/0007153];
W.~Taylor,
JHEP {\bf 0008}, 038 (2000)
[arXiv:hep-th/0008033];
H.~Hata and S.~Teraguchi,
JHEP {\bf 0105}, 045 (2001)
[arXiv:hep-th/0101162];
I.~Ellwood and W.~Taylor,
Phys.\ Lett.\ B {\bf 512}, 181 (2001)
[arXiv:hep-th/0103085];
L.~Rastelli, A.~Sen and B.~Zwiebach,
arXiv:hep-th/0012251.

\bibitem{cubic}
E.~Witten,
Nucl.\ Phys.\ B {\bf 268}, 253 (1986).
N.~Berkovits,
Nucl.\ Phys.\ B {\bf 450}, 90 (1995)
[Erratum-ibid.\ B {\bf 459}, 439 (1995)]
[arXiv:hep-th/9503099].

\bibitem{zwie}
J.~A.~Minahan and B.~Zwiebach,
JHEP {\bf 0009}, 029 (2000)
[arXiv:hep-th/0008231];
JHEP {\bf 0103}, 038 (2001)
[arXiv:hep-th/0009246];
JHEP {\bf 0102}, 034 (2001)
[arXiv:hep-th/0011226].

\bibitem{bsft}
E.~Witten,
Phys.\ Rev.\ D {\bf 46}, 5467 (1992)
[arXiv:hep-th/9208027];
Phys.\ Rev.\ D {\bf 47}, 3405 (1993)
[arXiv:hep-th/9210065];
S.~L.~Shatashvili,
Phys.\ Lett.\ B {\bf 311}, 83 (1993)
[arXiv:hep-th/9303143].


\bibitem{bsfttach}
A.~A.~Gerasimov and S.~L.~Shatashvili,
JHEP {\bf 0010}, 034 (2000)
[arXiv:hep-th/0009103];
D.~Kutasov, M.~Marino and G.~W.~Moore,
JHEP {\bf 0010}, 045 (2000)
[arXiv:hep-th/0009148];
arXiv:hep-th/0010108.

\bibitem{effective}
M.~R.~Garousi,
Nucl.\ Phys.\ B {\bf 584}, 284 (2000)
[arXiv:hep-th/0003122];
E.~A.~Bergshoeff, M.~de Roo, T.~C.~de Wit, E.~Eyras and S.~Panda,
JHEP {\bf 0005}, 009 (2000)
[arXiv:hep-th/0003221];
J.~Kluson,
Phys.\ Rev.\ D {\bf 62}, 126003 (2000)
[arXiv:hep-th/0004106].

\bibitem{0009061}
G.~W.~Gibbons, K.~Hori and P.~Yi,
``String fluid from unstable D-branes,''
Nucl.\ Phys.\ B {\bf 596}, 136 (2001)
[arXiv:hep-th/0009061].

\bibitem{0002223}
O.~Bergman, K.~Hori and P.~Yi,
``Confinement on the brane,''
Nucl.\ Phys.\ B {\bf 580}, 289 (2000)
[arXiv:hep-th/0002223];
J.~A.~Harvey, P.~Kraus and F.~Larsen,
JHEP {\bf 0012}, 024 (2000)
[arXiv:hep-th/0010060];
F.~Larsen,
Int.\ J.\ Mod.\ Phys.\ A {\bf 16}, 650 (2001)
[arXiv:hep-th/0010181].

\bibitem{older}
U.~Lindstrom and R.~von Unge,
Phys.\ Lett.\ B {\bf 403}, 233 (1997)
[arXiv:hep-th/9704051];
H.~Gustafsson and U.~Lindstrom,
Phys.\ Lett.\ B {\bf 440}, 43 (1998)
[arXiv:hep-th/9807064];
U.~Lindstrom, M.~Zabzine and A.~Zheltukhin,
JHEP {\bf 9912}, 016 (1999)
[arXiv:hep-th/9910159].

\bibitem{as}
A.~Sen,
JHEP {\bf 9910}, 008 (1999)
[arXiv:hep-th/9909062];
J.\ Math.\ Phys.\  {\bf 42}, 2844 (2001)
[arXiv:hep-th/0010240].

\bibitem{rrform}
E.~Witten,
JHEP {\bf 9812}, 019 (1998)
[arXiv:hep-th/9810188];
P.~Horava,
Adv.\ Theor.\ Math.\ Phys.\  {\bf 2}, 1373 (1999)
[arXiv:hep-th/9812135];
A.~Sen,
arXiv:hep-th/9904207.


\end{thebibliography}
\end{document}